\documentclass[aps,a4paper, preprint,floatfix,nofootinbib,amsmath,amssymb]{revtex4}
\usepackage{amstext,amssymb}
\usepackage{amsmath}
\usepackage{multirow}
\usepackage{graphicx}
\usepackage{dcolumn}
\usepackage[hyperfootnotes=false]{hyperref}
\usepackage{xspace}
\usepackage{braket}
\usepackage{color}
\usepackage{units}
\usepackage{comment}

\usepackage{slashed} % feynman slash notation via \slashed{p}
\usepackage{graphicx}
\usepackage{bm}% bold math
\newcommand {\ignore}[1]{}

\def\3331{$\mathrm{SU(3)_c \times SU(3)_L \times SU(3)_R \times U(1)_{X}}$}

\newcommand{\be}{\begin{equation}}
\newcommand{\ee}{\end{equation}}
\newcommand{\bea}{\begin{eqnarray}}
\newcommand{\eea}{\end{eqnarray}}
\newcommand{\nn}{\nonumber}
\begin{document}
\title{Majorana dark matter and neutrino mass with $S_3$ symmetry}

\author{Subhasmita Mishra}
\email{subhasmita.mishra92@gmail.com}
\affiliation{Department of Physics, IIT Hyderabad,
              Kandi-502285, India }                            

%%%%%%%%%%%%%%%%%%%%%%%%%%%%%%%%%%%%%%%%%%%%%%%%%%%%%%%%%%%%
\begin{abstract}
This model includes a minimal extension of the standard model with $S_3$ and $Z_2$ symmetries to explain neutrino masses and mixing along with the dark matter phenomenology. Neutrino phenomenology is explored, consistent with the $3 \sigma$ observation of oscillation parameters and a nonzero reactor mixing angle ($\theta_{13}$) is obtained. The $S_3$ singlet Majorana neutrino couples to the third generation of leptons, gives a correct relic density compatible with the 
Planck data. This model does not allow  tree level direct detection, therefore we discuss the loop level effective interaction  with the nucleus mediated by gauge boson. Also the constraints from the lepton flavor violating rare decay mode is commented. 
\end{abstract}
\maketitle
%%%%%%%%%%%%%%%%%%%%%%%%%%%%%%%%%%%%%%%%%%%%%%%%%%%%%%%%%%%%%%%%%%%%%%%%%%%%%%%%
\section{Introduction}
Albeit the tremendous success of standard model (SM), it fails to explain certain experimental evidences like neutrino masses, dark matter (DM), matter-anti matter asymmetry of the universe etc \cite{Ahmed:2003kj}-\cite{DiGangionbehalfoftheXENONCOLLABORATION:2018xht}. Hence to accommodate the explanation of these discrepancies, SM particle spectrum needs to be extended with extra symmetries. Moreover, the simplistic approach for explicit study of neutrino phenomenology with a compatibility of oscillation data, is to impose discrete symmetries in SM gauge group. This has been widely discussed in the literature \cite{King:2015bja}-\cite{DeRujula:1977dmn}. $A_4$ and $S_3$ symmetries are very well approached  by the phenomenologist to address the neutrino issues along with various cosmological problems like particle candidates of DM and baryon asymmetry of the universe etc. Imposition of discrete symmetries restrict the Yukawa interaction and hence provides a specific structure to the neutrino mass matrix with an interesting phenomenology. Therefore helps in exploring the neutrino masses and mixing, evident from various experiments. 

  Apart from the neutrino issues, the mischievous existence of dark matter(DM) is evidenced from various observations like galaxy rotation curves, gravitational lensing etc. The stable and non-baryonic nature of this mysterious matter can be inferred from the Cosmic Microwave Background radiation(CMB) and large scale structure of the universe. On the other hand, particle candidate of the DM is not yet known, neither the detection of DM by various experiments has achieved a huge success in few decades. But there is definitely a substantial improvement in collection of data by different pro active experiments like LUX,XENON 1T, PandaX, LEP II, ATLAS, CMS for direct and IceCube, Fermi-LAT, AMS etc for indirect searches in the present era \cite{Akerib:2015rjg}-\cite{Aprile:2017iyp}. The WIMP miracle has given a new direction to the DM searches, which is proven out by certain experimental upper bounds on the DM-nucleon interaction cross section \cite{Bertone:2004pz}-\cite{Agrawal:2010fh}.

   However, extension of SM with the simplest permutation symmetry $S_3$ is well described in various literature \cite{Meloni:2010aw}-\cite{Mondragon:2006hi}. The two dimensional irreducible representations makes the phenomenological aspects more easier and interesting. This symmetry is widely used for the study of neutrino phenomenology and also leptogenesis within type I and type II seesaw framework. Very few literature devoted to extensive study of DM within the framework of $S_3$ symmetry \cite{Espinoza:2018itz}. In this article,  SM is extended with $S_3$ symmetry to study the neutrino masses and mixing compatible with $3 \sigma$ observation of neutrino oscillation data. Along with the SM particle spectrum, three extra right-handed Majorana fermions and two Higgs doublets are included. $Z_2$ symmetry is imposed to ensure the stability of Majorana DM.
      
      The article is structured as, section II includes the detail description of the particle content of the model. In section III, the neutrino masses and mixing are discussed. Section IV includes the detail study of DM phenomenology. In section V, I commented on the lepton flavor violating rare decay constraints and finally summarize the work in section VI.

\section{The model framework}

Here, we discuss the particle content and corresponding group charges of the lepton sector. The extension of the SM ($SU(3)\times SU(2)_L\times U(1)_Y)$) with the simplest non-abelian discrete flavor symmetry, $S_3$, and the abelian symmetry $Z_2$ is discussed in addition to three right handed neutrinos ($N_{(1,2,3)R}$) and two Higgs doublets to explain the neutrino phenomenology, dark matter and lepton flavor violating decays. Particle spectrum of this framework transform as irreducible representations of $S_3$ group. First two generations of the left and right fermions of the model transform as a doublet under $S_3$, where, the third generations remain as singlets. Three Higgs doublets corresponding to three generations, also transform in a similar way. Third generation right-handed neutrino and Higgs are imposed to be odd under $Z_2$, where, the lightest mass eigenstate could be suitable DM candidate.
\begin{table}[h]
\begin{center}
\begin{tabular}{| c | c | c | c |}
\hline
~Particles~ & ~SM - Group ~ &~ $S_3$ ~ &~ $Z_2$ ~\\
\hline
\hline
 $L_e, L_\mu$ & ($1,2,-1$)  & $2$ & $+1$\\
  $L_\tau$ & ($1,2,-1$) & $1$ & $+1$ \\
  $E_{1R}, E_{2R}$ & ($1,1,-2$) & $2$ & $+1$\\
  $E_{3R}$ & ($1,1,-2$) & $1$ & $+1$  \\
   $N_{1R}, N_{2R}$ & ($1,1,0$) & $2$ & $+1$\\
   $N_{3R}$ & ($1,1,0$) & $1$ & $-1$\\ 
   $H_1, H_2$ & ($0,2,1$) & $2$ & $+1$\\
   $H_3$ & ($0,2,1$) & $1$ & $-1$ \\
 \hline
\end{tabular}
\end{center}
\caption{Particle spectrum and group charge under SM-gauge group and $ S_3 \otimes Z_2$.}\label{particles}
\end{table}
\subsection{Scalar Potential and symmetry breaking}
As this model retains three electroweak Higgs doublets, one can write the interaction potential as following \cite{Emmanuel-Costa:2016vej}-\cite{Borah:2017dfn} 
\begin{eqnarray}
V&=&\mu^2_1({H_2}^{\dagger}H_2+{H_1}^{\dagger}H_1)+\mu^2_3 {H_3}^{\dagger}H_3+\lambda_1({H_2}^{\dagger}H_2+{H_1}^{\dagger}H_1)^2 \nn \\
&&+\lambda_2({H_1}^\dagger H_2-{H_2}^\dagger H_1)^2+\lambda_3[({H_1}^\dagger H_1-{H_2}^\dagger H_2)^2+({H_1}^\dagger H_2+{H_2}^\dagger H_1)^2] \nn \\
&&+\lambda_4[({H_3}^\dagger H_1)({H_1}^\dagger H_2+{H_2}^\dagger H_1)+({H_3}^\dagger H_2)({H_1}^\dagger H_1-{H_2}^\dagger H_2)+h.c] \nn \\
&&+\lambda_5[({H_3}^\dagger H_3)({H_1}^\dagger H_1+{H_2}^\dagger H_2)]+\lambda_6[({H_3}^\dagger H_1)({H_1}^\dagger H_3)+({H_3}^\dagger H_2)({H_2}^\dagger H_3)] \nn \\
&&+\lambda_7[({H_3}^\dagger H_1)({H_3}^\dagger H_1)+({H_3}^\dagger H_2)({H_3}^\dagger H_2)+h.c]+\lambda_8({H_3}^\dagger H_3)^2 \nn \\
&&+\mu^2_{SB1}({H_1}^\dagger H_2+h.c). 
\end{eqnarray}
With consideration of the symmetry breaking pattern, the doublets being charged under the electroweak symmetry, contribute to the breaking of $\rm SU(2)_L\otimes U(1)_Y$. The minimization conditions are given by $\frac{\partial V}{\partial v_1}=0$ and $\frac{\partial V}{\partial v_2}=0$ where, $\langle H_1 \rangle=\begin{pmatrix} 0 && v_1 \end{pmatrix}^T$,  $\langle H_2 \rangle=\begin{pmatrix} 0 && v_2 \end{pmatrix}^T$. The third Higgs being odd under $Z_2$, does not acquire any vacuum expectation value (VEV). Multi Higgs models give rise to tree level flavor changing neutral current, which can be avoided by a heavy Higgs mass of order TeV. This scale can not be achieved by the electroweak symmetry breaking. Therefore, the explicit symmetry breaking term is introduced in the potential, which leads to a mixing between the first two Higgs doublets. Apart from the symmetry breaking, the stability conditions of the scalar potential by using the co-positivity criteria \cite{Kannike:2012pe} are given below  
\begin{eqnarray}
&& \lambda_1+\lambda_3\geq 0, \hspace{2mm} \lambda_8\geq 0, \nn \\
&& \lambda_5+\lambda_6+|\lambda_7|+\sqrt{\lambda_8(\lambda_3+\lambda_1)}\geq 0, \nn \\
&& 3(\lambda_1+\lambda_3)\sqrt{\lambda_8}+2(\lambda_5+\lambda_6+|\lambda_7|)\sqrt{\lambda_1+\lambda_3}\geq 0, \nn  \\
% && 4(\lambda_1+\lambda_3)\sqrt{\lambda_8}+(2+\sqrt{6})(\lambda_5+\lambda_6+|\lambda_7|)\sqrt{\lambda_1+\lambda_3}\geq0\\
&& 2(\lambda_5+\lambda_6+|\lambda_7|)^2-3\lambda_8(\lambda_1+\lambda_3)\geq 0.
\label{stability}
\end{eqnarray}
\subsection{Higgs masses and mixing}
Here, we can write the mass basis of the first two Higgs doublets by orthogonal rotation of flavor states as following \cite{Araki:2005ec} ,
\begin{eqnarray}
\begin{pmatrix}
H_1\\
H_2\\
\end{pmatrix}=\begin{pmatrix}
\cos{\beta} & \sin{\beta}\\
-\sin{\beta} & \cos{\beta}\\
\end{pmatrix} \begin{pmatrix}
H_L\\
H_H\\
\end{pmatrix},
\label{higgsmixing}
\end{eqnarray}
here, $H_L=H_1 \cos{\beta}-H_2 \sin{\beta}$ and $H_H=H_1 \sin{\beta} + H_2 \cos{\beta} $, and $\beta$ is the Higgs mixing angle. The Higgs doublets in the mass eigenstate are written in an electroweak doublet form as  
\begin{eqnarray}
H_L=\begin{pmatrix}
0\\
h^0_L+v
\end{pmatrix},~~~  H_H=\begin{pmatrix}
h^+_H\\
h^0_H+ia_H\\
\end{pmatrix},~~~  H_3=\begin{pmatrix}
h^+_3\\
h^0_3+ia_3
\end{pmatrix}.
\end{eqnarray}
Here, $H_L$ is the SM like Higgs and the alignment of vacuum expectation value leads to $v=\sqrt{{v_1}^2+{v_2}^2}=246$ GeV and $\tan{\beta}=\frac{v_2}{v_1}$. Charged and CP odd components of $H_L$ will be absorbed by the SM gauge bosons to acquire mass in unitary gauge conditions. And rest of the Higgs doublets will have massive CP odd, charged and neutral scalar fields, which will contribute to the phenomenology of this model. As previously mentioned, the extra Higgs fields should be much heavier to suppress the tree level FCNC, which can achieved by the finetunning of the explicit symmetry breaking parameter. The third Higgs doublet being odd under $Z_2$, would not have direct interaction with the three generations simultaneously and hence does not contribute to the FCNCs. The masses of the heavy Higgs will be of order TeV, which can be approximated as $M^2_{h_H}\approx M^2_{h^+_{H}}\approx M^2_{a_H}\approx \mathcal{O}(\mu^2_{SB} \sin{2\beta})$ and the SM like Higgs will have a squared mass of $\mathcal{O}(v^2)$. The mass of the inert Higgs is adjusted to study the DM phenomenology in later sections.
\subsection{Interaction Lagrangian and Leptonic mixing matrices }
To discuss the model phenomenology, we started with the particle content and their corresponding group charges with respect to the SM and $S_3\times Z_2$ symmetries in Table I. The  Yukawa interaction Lagrangian for charged and neutral sectors, that involves the new scalars and fermions in the current framework is given by \cite{Mishra:2019sye},\cite{Araki:2005ec}, 
\begin{eqnarray}
\mathcal{L}_{Mass}&=&-y_1 \left[ \overline{L_e} \tilde{H_2} N_{1R} +\overline{L_\mu} \tilde{H_1} N_{1R}+\overline{L_e} 
\tilde{H_1} N_{2R}-\overline{L_\mu} \tilde{H_2} N_{2R} \right] \nonumber\\
               && - y_3\left[ \, \overline{L_\tau}  \,  \,  
         \tilde{H_1} N_{1R}+\, \overline{L_\tau}  \, \,  
         \tilde{H_2}  N_{2R}\right]- y_5\,\left[\overline{L_\tau} \,  \, 
         \tilde{H_3} N_{3R} \right] \nonumber \\
&&- y_{l2} \left[\overline{L_e} {H_2} E_{1R} +\overline{L_\mu} {H_1} E_{1R}+
        \overline{L_e}  {H_1} E_{2R}-\overline{L_\mu} {H_2} E_{2R}\right] \nonumber \\
&&- y_{l4}\left[ \, \overline{L_\tau}  \, \,  {H_1}  E_{1R}+\overline{L_\tau}  \,  \,  {H_2} E_{2R}\right]- 
y_{l5} \, [\overline{L_e} \, \,{H_1}  E_{3R}+\overline{L_\mu} \,  \, {H_2} E_{3R}]\nonumber\\
&& -\frac{1}{2} \sum_{i=1,2}\overline{N}_\text{iR}^\text{c} M_\text{iR} N_\text{iR} -\frac{1}{2}\overline{N}_\text{3R}^\text{c} M_\text{3R} N_\text{3R}+
	\rm{~h.c}.\label{Yukawa}
\end{eqnarray}
%\subsection{Neutrino Phenomenology}
Before discussing the neutrino mixing, we start with the full mass matrix of well known type I seesaw in the basis $\tilde{N} = (\nu_\text{L}^\text{c}, ~N_\text{R})^\text{T}$, which is given by
%================================================
\begin{equation}
	\mathcal{M} = 
	\begin{pmatrix}
		0 & m_\text{D} \\
		m_\text{D}^\text{T} & M_R  \\
	\end{pmatrix}.	\nonumber	
\end{equation}
Furthermore, in this context, we consider the light neutrino mass formula well described by the known type I seesaw mechanism as,
\begin{align}
	m_\nu &= M_\text{D} M_{R}^{-1} \left( M_\text{D}  \right)^T. \label{type I}
\end{align}
From the interaction Lagrangian in Eq.\eqref{Yukawa}, we can write the flavor structure of Dirac mass matrix for neutral and charged leptons as following.
\begin{eqnarray}
M_D=\begin{pmatrix}
   y_1 v_2 & y_1 v_1      & 0 \\
   y_1 v_1 & -y_1 v_2 & 0 \\
   y_3 v_1      & y_3 v_2       & 0 \\
   \end{pmatrix},\hspace{2mm}
M_l=\begin{pmatrix}
   y_{l2} v_2 & y_{l2} v_1       & y_{l5} v_1 \\
   y_{l2} v_1       & -y_{l2} v_2 & y_{l5} v_2 \\
   y_{l4} v_1       & y_{l4} v_2       & 0\\
   \end{pmatrix}.\label{massmatrices} 
\end{eqnarray}
\section{Neutrino masses and mixing}
In order to study the neutrino oscillation phenomenology, we consider the results of  oscillation parameters by using the $3 \sigma$ observations, which are represented in Table II. The standard neutrino mixing matrix is provided below
\begin{eqnarray}
U_{PMNS}= \begin{pmatrix}
c_{13} c_{12}        &c_{13} s_{12}    & s_{13} e^{-i \delta}\\
-c_{23} s_{12}-c_{12} s_{13} s_{23}e^{i \delta} &c_{12}c_{23}-s_{12}s_{13}s_{23}e^{i \delta}   &c_{13} s_{23}\\
s_{12} s_{23}-c_{12}s_{13}c_{23}e^{i \delta}  & -c_{12}s_{23}-c_{23}s_{13}s_{12}e^{i \delta}   &c_{23} c_{13}\\
\end{pmatrix}  \begin{pmatrix}
1     &0  &0\\
0    & e^{i \alpha}  &0\\
0    &0        & e^{i \beta}\\
\end{pmatrix}.
\end{eqnarray}

Here, $c_{ij}= \cos{\theta}_{ij}$ and $s_{ij}= \sin \theta_{ij}$ are the mixing angles and $\delta,\alpha$ and $\beta $ are the Dirac and Majorana phases respectively.
\begin{table}[htbp]
\centering
{
\renewcommand{\arraystretch}{1.2}
\catcode`?=\active \def?{\hphantom{0}}
\begin{minipage}{\linewidth}
\begin{tabular}{|l|c|c|c|}
\hline
parameter & best fit $\pm$ $1\sigma$ &  2$\sigma$ range& 3$\sigma$ range
\\
\hline\hline
$\Delta m^2_{21}\: [10^{-5}~ {\rm eV}^2]$ & 7.56$\pm$0.19  & 7.20--7.95 & 7.05--8.14 \\
\hline
$|\Delta m^2_{31}|\: [10^{-3} ~{\rm eV}^2]$ (NO) &  2.55$\pm$0.04 &  2.47--2.63 &  2.43--2.67\\
$|\Delta m^2_{31}|\: [10^{-3}~ {\rm eV}^2]$ (IO)&  2.47$^{+0.04}_{-0.05}$ &  2.39--2.55 &  2.34--2.59 \\
\hline
$\sin^2\theta_{12} / 10^{-1}$ & 3.21$^{+0.18}_{-0.16}$ & 2.89--3.59 & 2.73--3.79\\
\hline
  $\sin^2\theta_{23} / 10^{-1}$ (NO)
	  &	4.30$^{+0.20}_{-0.18}$ 
	& 3.98--4.78 \& 5.60--6.17 & 3.84--6.35 \\
  $\sin^2\theta_{23} / 10^{-1}$ (IO)
	  & 5.98$^{+0.17}_{-0.15}$ 
	& 4.09--4.42 \& 5.61--6.27 & 3.89--4.88 \& 5.22--6.41 \\
\hline %%	
$\sin^2\theta_{13} / 10^{-2}$ (NO) & 2.155$^{+0.090}_{-0.075}$ &  1.98--2.31 & 1.89--2.39 \\
$\sin^2\theta_{13} / 10^{-2}$ (IO) & 2.155$^{+0.076}_{-0.092}$ & 1.98--2.31 & 1.90--2.39 \\
    \hline
  \end{tabular}
  \caption{ \label{tab:sum-2017} 
   The experimental values of neutrino oscillation parameters for $1\sigma$, $2\sigma$ and $3\sigma$ range \cite{deSalas:2017kay}.}
    \end{minipage}
  }
\end{table} 

The above experimental data gives rise to two scenarios of neutrino masses: the Normal Hierarchy: $m_1 << m_2 <<m_3$ and 
the Inverted Hierarchy: $m_3 << m_1 <<m_2$. In Table II $\Delta m_{21}^2$ is known as the solar mass square difference and $\Delta m_{23}^2$ and 
$\Delta m_{31}^2$ are the atmospheric mass squared differences for different hierarchies quoted. 
\subsection{Diagonalization of neutrino mass matrix}
The Dirac mass matrix in Eq.\eqref{massmatrices} is modified after the rotation and VEV alignment of the Higgs fields. Therefore one can write the type I seesaw neutrino mass in Eq.\eqref{type I} from the modified Dirac mass as following
\begin{equation}
M_D=\begin{pmatrix}
   y_1 v\sin{\beta} & y_1 v\cos{\beta}      & 0 \\
   y_1 v\cos{\beta} & -y_1 v\sin{\beta} & 0 \\
   y_3 v\cos{\beta}      & y_3 v\sin{\beta}  & 0 \\
   \end{pmatrix},~~\mathcal{M_\nu}=\begin{pmatrix}
\frac{y^2_1 v^2}{M_{1R}} && 0 && \frac{y_1 y_3 v^2 \sin{2\beta}}{M_{1R}}\\
0 && \frac{y^2_1 v^2}{M_{1R}} && \frac{y_1 y_3 v^2 \cos{2\beta}}{M_{1R}}\\
\frac{y_1 y_3 v^2 \sin{2\beta}}{M_{1R}} && \frac{y_1 y_3 v^2 \cos{2\beta}}{M_{1R}} && \frac{y^2_3 v^2}{M_{1R}} 
\end{pmatrix}.
\end{equation}
The above mass matrix can be diagonalized by an unitary eigenvector matrix with mass eigenvalues are given by
\begin{eqnarray}
&& m_{\nu_1}=\frac{y^2_1 v^2}{M_{1R}},~~ m_{\nu_2}=\frac{(y^2_1+y^2_3) v^2}{M_{1R}},~~ m_{\nu_3}=0.
\end{eqnarray}
\begin{figure}
\includegraphics[height=36mm,width=58mm]{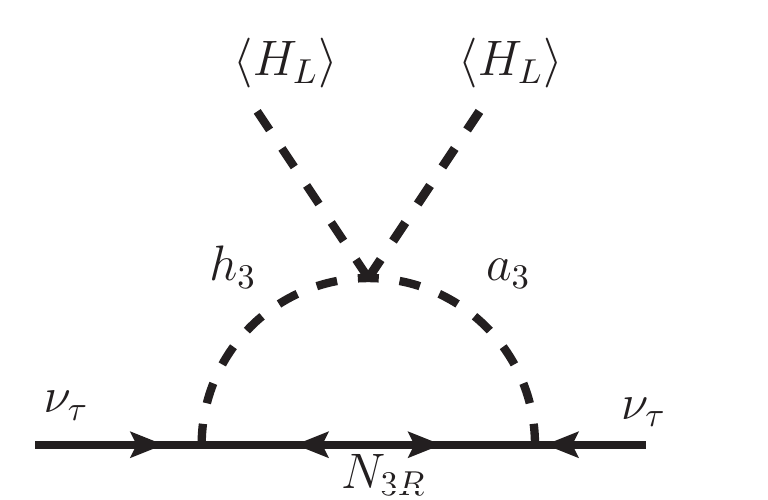}
\caption{Radiatively generated third generation neutrino mass.}\label{radmass}
\end{figure}
 We can have the third generation neutrino to have a vanishing mass eigenvalue after diagonalization. Even though two mass parameters of the neutrinos are enough to explain the neutrino oscillation data, we can still allow a Majorana mass term for the third neutrino by radiative correction, shown in Fig.\ref{radmass}. The expression for the radiatively generated neutrino mass is given by \cite{Ma:2006km}-\cite{Borah:2018uci} 
 \begin{eqnarray}
 m_{{\nu}_3}=\frac{{y^2_5}}{16 \pi^2} M_{3R}\left[\frac{{M^2_{h^0_3}}}{M^2_{h^0_3}-{M^2_{3R}}} {\rm ln}\left(\frac{{M^2_{h^0_3}}}{{M^2_{3R}}}\right)-\frac{{M^2_{a_3}}}{{M^2_{a_3}}-{M^2_{3R}}} {\rm ln}\left(\frac{{M^2_{a_3}}}{{M^2_{3R}}}\right)\right].%\includegraphics[width=40mm,height=30mm]{Neutrino_mass_3.eps}
 \end{eqnarray}
Here, $M_{h^0_3}$ and $M_{a_3}$ are the masses of the CP even and odd component of the inert Higgs. Under the approximation of $M^2_{3R} \approx \frac{1}{2}(M^2_{h^0_3}+M^2_{a_3})$, the above expression reduced to a simplified form,
\begin{equation}
m_{{\nu}_3} \approx  \frac{\lambda_7 v^2}{16 \pi^2}\frac{y^2_5 }{M_{3R}}\,.
\label{numass2}
\end{equation}
The mass splitting between the CP odd and CP even component is given by
\be
M^2_{h^0_3}-M^2_{a_3} = \lambda_7 v^2\,.
\ee
The mixing matrix that diagonalizes the Majorana mass matrix of the neutrino is constructed from the eigenvectors of the neutrino mass matrix and is given as following
\begin{eqnarray}
U_\nu=\begin{pmatrix}
-\cos{\beta} && \frac{y_1}{\sqrt{y^2_1+y^2_3}}\sin{2\beta} && -\frac{y_3}{\sqrt{y^2_1+y^2_3}}\sin{2\beta}\\
\sin{2\beta} && \frac{y_1}{\sqrt{y^2_1+y^2_3}}\cos{2\beta} && -\frac{y_3}{\sqrt{y^2_1+y^2_3}} \cos{2\beta}\\
0 && \frac{y_3}{\sqrt{y^2_1+y^2_3}} && \frac{y_1}{\sqrt{y^2_1+y^2_3}} 
\end{pmatrix}.
\end{eqnarray}
The squared charged lepton mass matrix can be diagonalized by unitary transformation as $U_{eL} M_l {M_l}^\dagger {U_{eL}}^\dagger={\rm Diag}(m^2_e,\hspace{2mm}m^2_{\mu},\hspace{2mm} m^2_\tau)$. The mixing matrix for the squared charged lepton masses can be obtained by solving the characteristic equation \cite{Mondragon:2006hi}, 
\begin{eqnarray}
U_{el}=\begin{pmatrix}
   \frac{x}{\sqrt{2(1-x^2)}} && \frac{1}{\sqrt{2(1+x^2)}} && \frac{1}{\sqrt{2(1+\sqrt{z})}}\\
  \frac{-x}{\sqrt{2(1-x^2)}}  &&\frac{-1}{\sqrt{2(1+x^2)}}  && \frac{1}{\sqrt{2(1+\sqrt{z})}}\\
   \frac{\sqrt{1-2x^2}}{\sqrt{1-x^2}} && \frac{x}{\sqrt{1+x^2}} && \frac{\sqrt{z}}{\sqrt{(1+\sqrt{z})}}
   \end{pmatrix}.
   \label{lepton mixing matrix}
\end{eqnarray} 
Here, $x=\frac{m_e}{m_\mu}$, and $z= \frac{m_e^2m_\mu^2}{m_\tau^4}$.
\begin{figure}
\includegraphics[height=53mm,width=72mm]{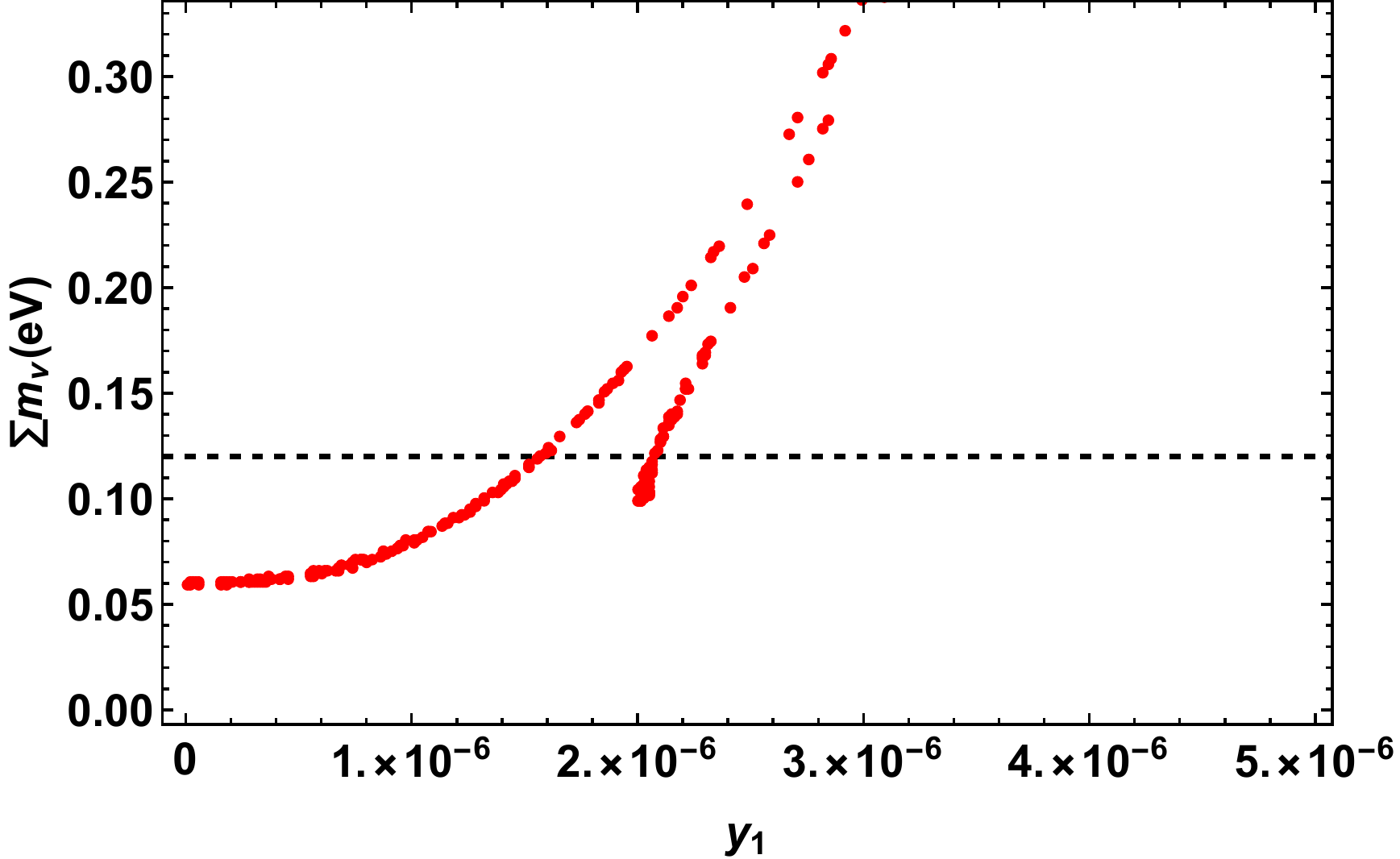}
\includegraphics[height=53mm,width=72mm]{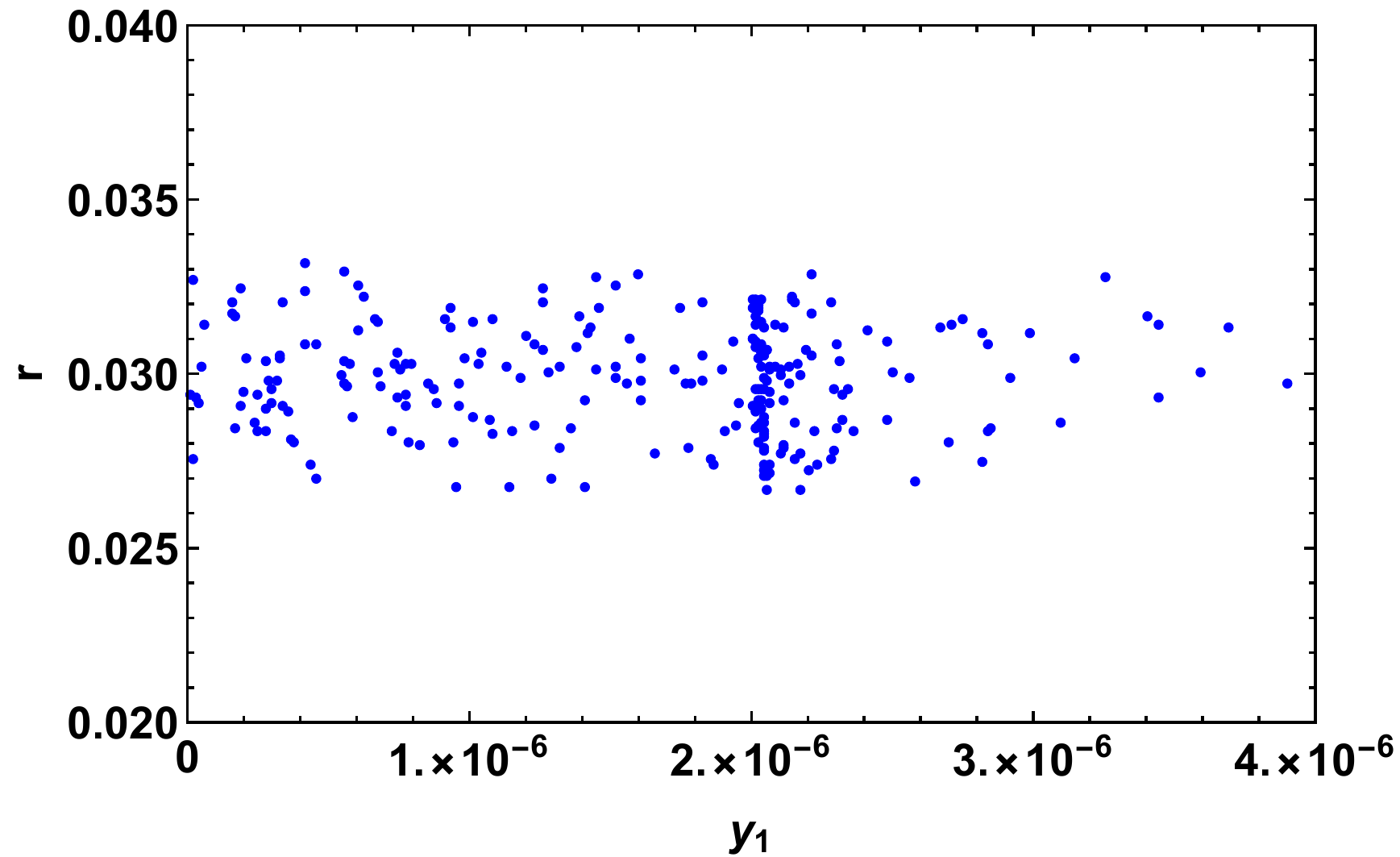}
\caption{Variation of the Yukawa coupling ($y_1$) with the sum of neutrino mass is displayed in the left panel and the the right panel shows the variation of $y_1$ with the ratio of solar to atmospheric mass squared differences(r).}\label{nu1}
\end{figure}
Therefore the standard $U_{\rm PMNS}$ mixing matrix for this model can be parameterized as $U_{\rm PMNS}=U^\dagger_{eL} U_\nu$.
\begin{eqnarray}
U_{\rm PMNS}=\begin{pmatrix}
 -\frac{x (\cos {2 \beta }+\sin {2 \beta })}{\sqrt{2-2 x^2}} & \frac{x (\sin {2 \beta }-\cos {2 \beta }) y_1+\sqrt{2-4 x^2} y_3}{\sqrt{2-2 x^2} \sqrt{y_1^2+y_3^2}} & \frac{\sqrt{2-4 x^2} y_1+x (\cos {2 \beta }-\sin {2 \beta }) y_3}{\sqrt{2-2 x^2} \sqrt{y_1^2+y_3^2}} \\
 -\frac{\cos {2 \beta }+\sin {2 \beta }}{\sqrt{2} \sqrt{x^2+1}} & \frac{(\sin {2 \beta }-\cos {2 \beta }) \sqrt{2} y_1+2 x y_3}{2 \sqrt{x^2+1} \sqrt{y_1^2+y_3^2}} & \frac{2 x y_1+(\cos {2 \beta }-\sin {2 \beta }) \sqrt{2} y_3}{2 \sqrt{x^2+1} \sqrt{y_1^2+y_3^2}} \\
 \frac{\sin {2 \beta }-\cos {2 \beta }}{\sqrt{2} \sqrt{\sqrt{z}+1}} & \frac{(\cos {2 \beta }+\sin {2 \beta }) \sqrt{2} y_1+2 \sqrt{z} y_3}{2 \sqrt{\sqrt{z}+1} \sqrt{y_1^2+y_3^2}} & \frac{2 \sqrt{z} y_1-\sqrt{2} (\cos {2 \beta }+\sin {2 \beta }) y_3}{2 \sqrt{\sqrt{z}+1} \sqrt{y_1^2+y_3^2}} \\
 \end{pmatrix}.\label{thupmns}
\end{eqnarray}
The neutrino mixing angles can be found from the above mixing matrix by comparing with the standard $U_{\rm PMNS}$ matrix, which are given as following
{\small{
\begin{eqnarray}
&& \sin{\theta_{13}}=|\frac{\sqrt{2-4 x^2} y_1+x (\cos {2 \beta }-\sin {2 \beta }) y_3}{\sqrt{2-2 x^2} \sqrt{y_1^2+y_3^2}}|,\\
&& \tan{\theta_{12}}=|-\frac{x (-\cos{2 \beta} + \sin{2 \beta})y_1 + \sqrt{(2 - 4 x^2)} y_3}{x(\cos{2\beta} + \sin{2 \beta}) \sqrt{y^2_1+y^2_3}}|,\\
&& \tan{\theta_{23}}=|\frac{\sqrt{(1+\sqrt{z})}(2xy_1+\sqrt{2}(\cos{2 \beta}-\sin{2 \beta})y_3)}{\sqrt{(1+x^2)}(2\sqrt{z}y_1+\sqrt{2}(\cos{2 \beta}+\sin{2 \beta})y_3)}|.\label{angles}
\end{eqnarray}}}
 As per the mentioned parameter entries of the mixing matrix in Eq.\eqref{thupmns}, one can infer a nonzero value of reactor mixing angle ($\theta_{13}$) provided in  Eq.\eqref{angles}. \\
\begin{figure}
\includegraphics[height=53mm,width=72mm]{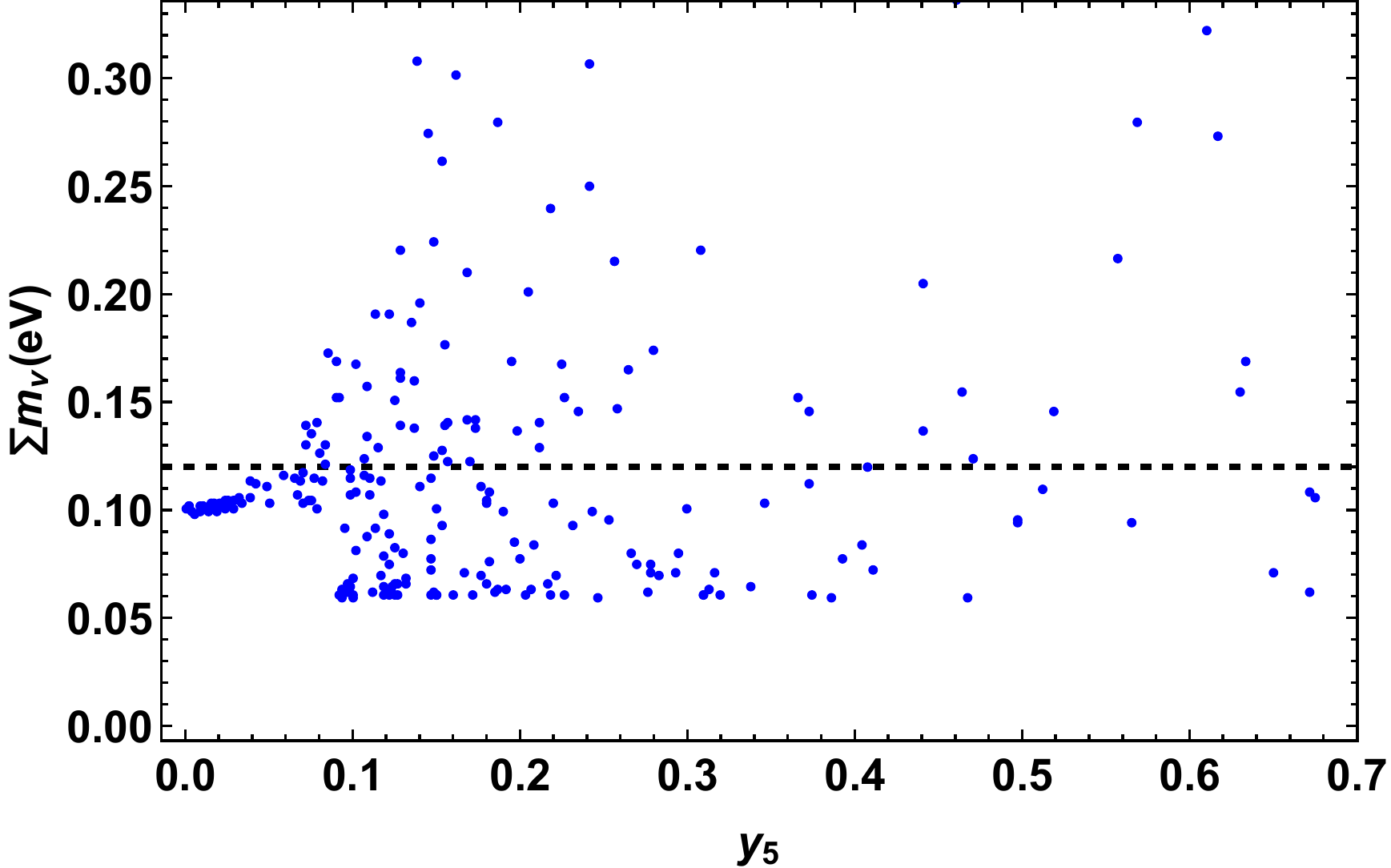}
\includegraphics[height=53mm,width=72mm]{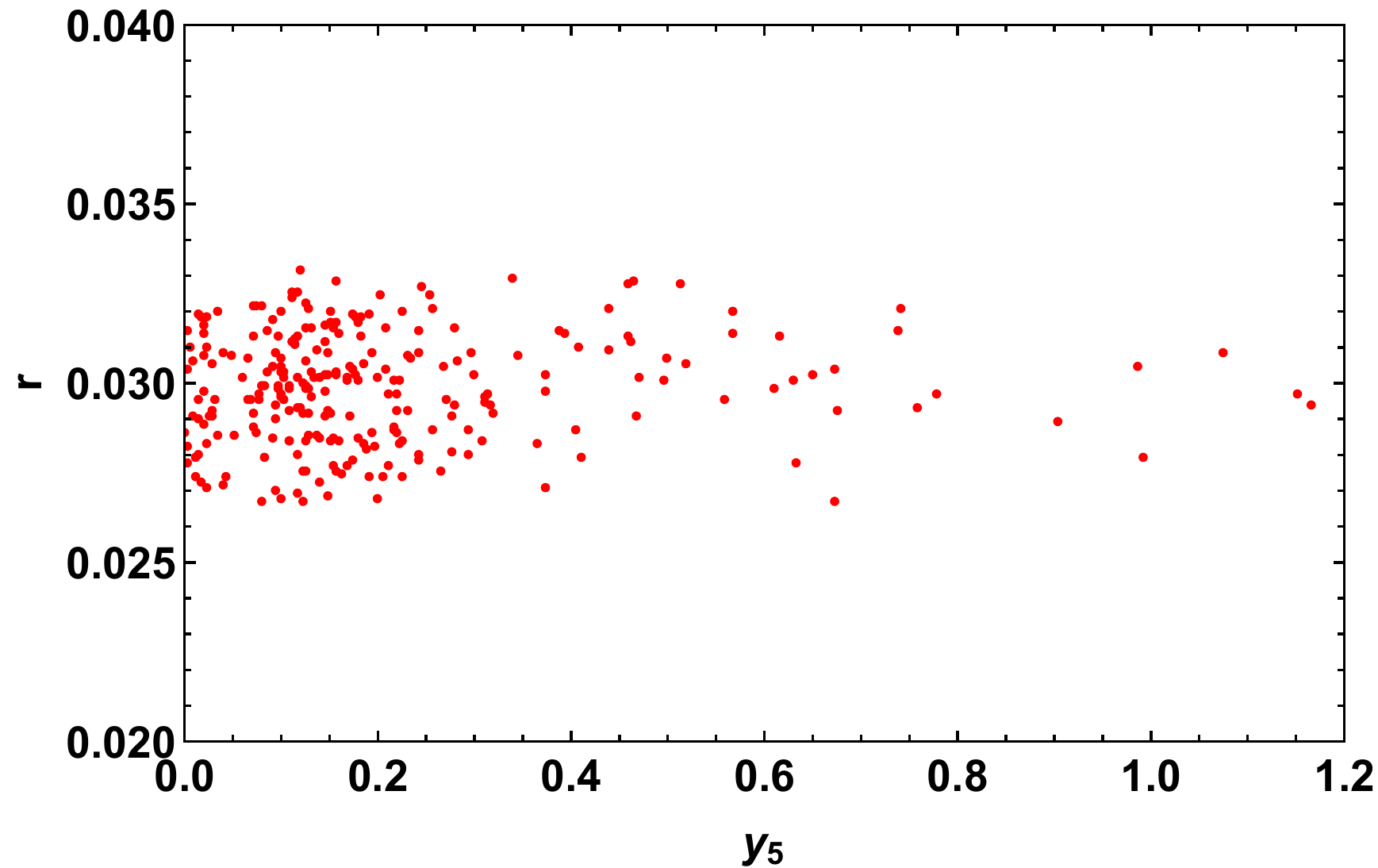}
\caption{Variation of $y_5$ with sum of the active neutrino masses is displayed in the left panel and variation with the ratio of solar to atmospheric mass squared differences(r) is shown in the right panel.}\label{nu2}
\end{figure}
\begin{figure}
\includegraphics[height=53mm,width=72mm]{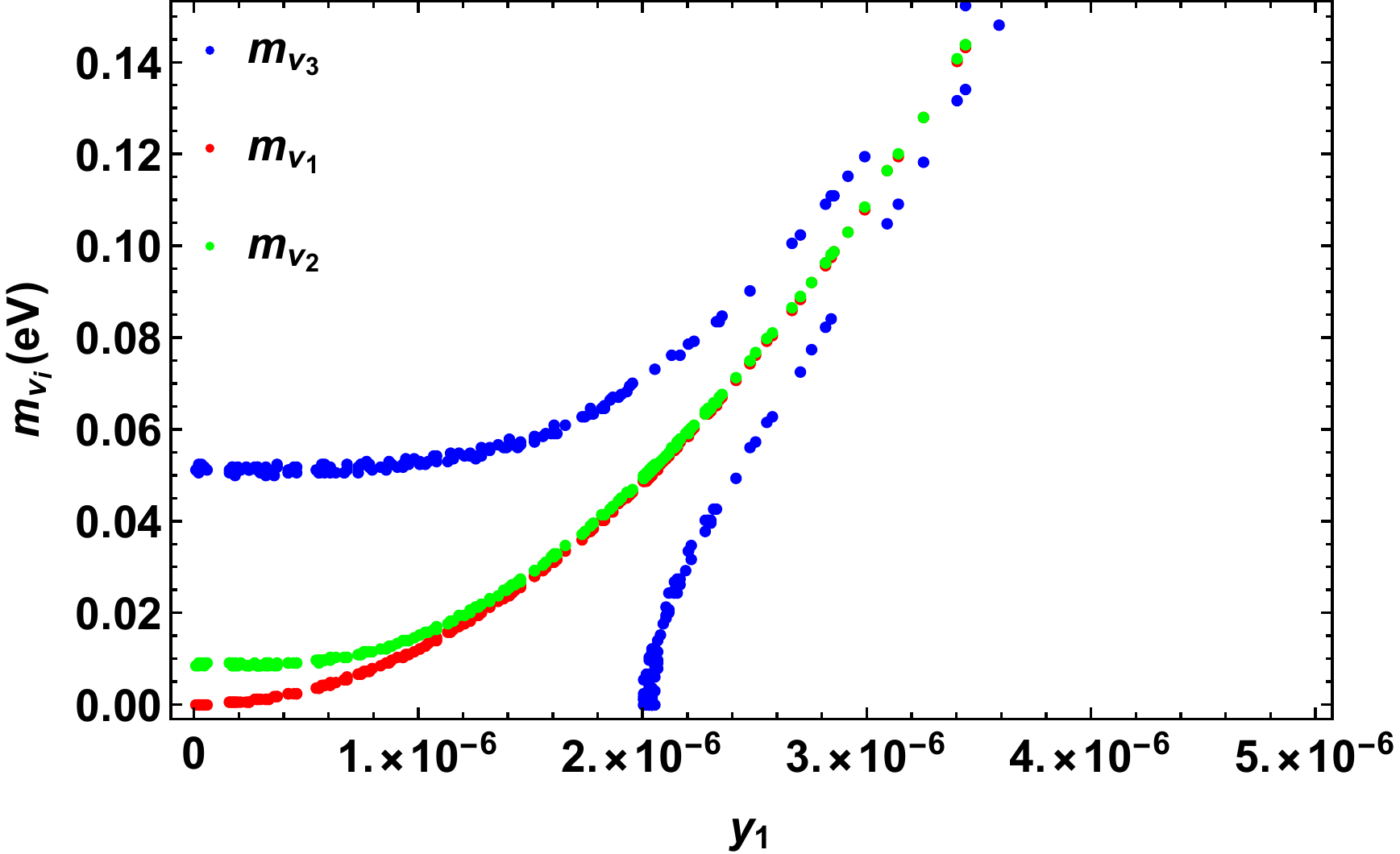}
\includegraphics[height=53mm,width=72mm]{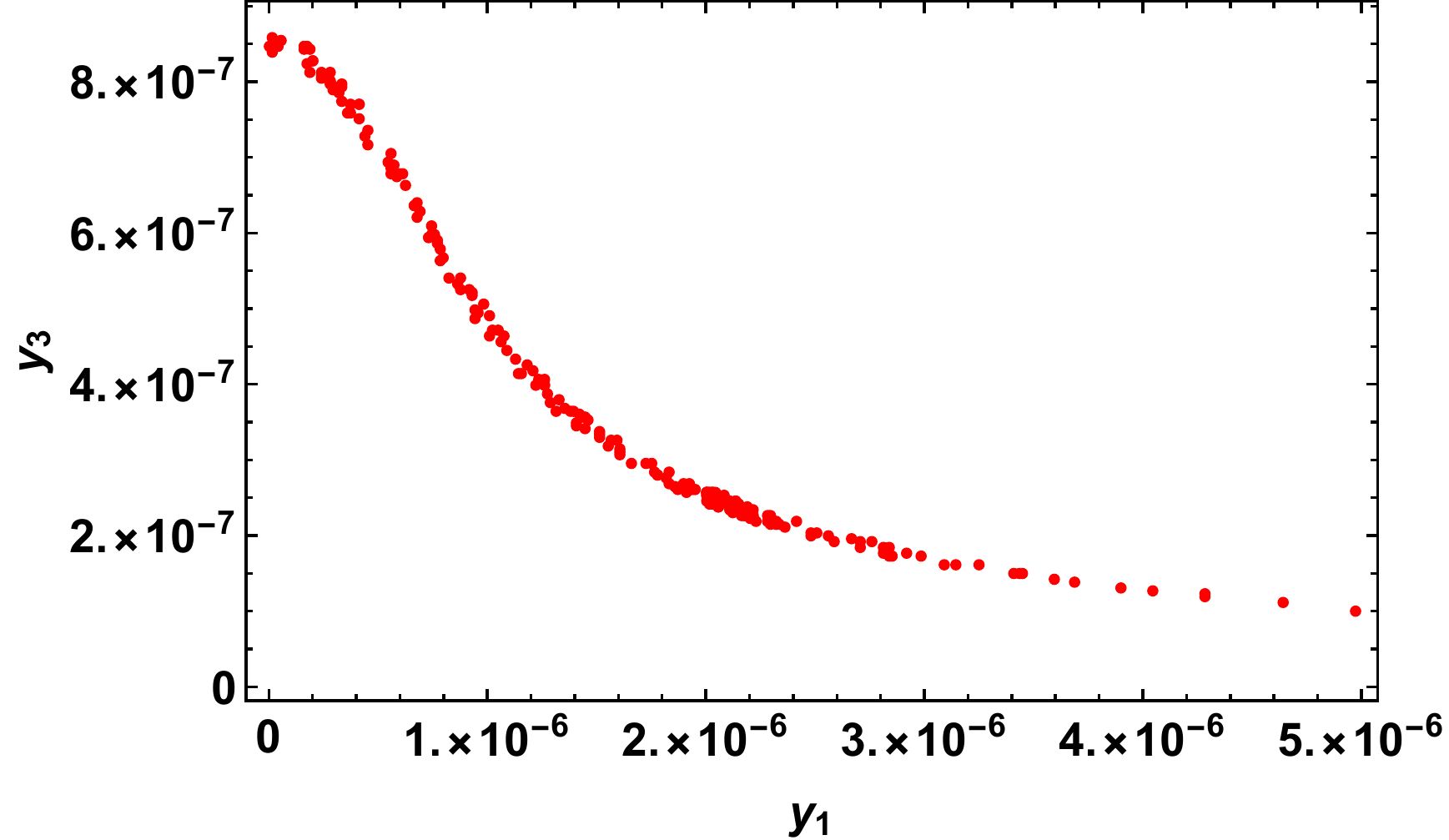}
\caption{Left panel shows the variation of $y_1$ with three active neutrino masses and the right panel displays a correlation between $y_1$ and $y_3$.}\label{nu3}
\end{figure}
      We discuss the dependence of various Yukawa coupling, which are consistent with the $3\sigma$ estimation of current neutrino oscillation data. We found from Fig.\ref{nu1}, that the values of Yukawa coupling greater than $2 \times 10^{-6}$ are excluded by the cosmological observation of total neutrino mass \cite{Aghanim:2018eyx}. The third generation neutrino acquires a radiative mass from the Yukawa coupling $y_5$, which is required  to be order of 1 to satisfy the correct DM relic and also lie within the bounds of neutrino data, presented in Fig.\ref{nu2}. Fig.\ref{nu3} displays a direct correlation of the Yukawa coupling $y_1$ with the active neutrino masses and the coupling $y_3$ consistent with the experimental observations.
\section{Dark matter}
This model allows the lightest right handed neutrino to be a dark matter candidate, which is stabilized by the $Z_2$ symmetry. Here the dark matter is allowed to have only scalar and lepton mediated t-channel annihilation process, which contribute to the relic density.
The relic abundance of $\rm DM$, can be obtained from the solution of the Boltzmann equation
\begin{equation}
\frac{dn_{\rm DM}}{dt}+3Hn_{\rm DM} = -\langle \sigma v \rangle (n^2_{\rm DM} -(n^{\rm eq}_{\rm DM})^2),
\end{equation}
here, $n^{\rm eq}_{\rm DM}$ is the equilibrium number density of $\rm DM$, $H$ stands for the Hubble expansion rate of the Universe. $ \langle \sigma v \rangle $ is the thermally averaged annihilation cross section of $\rm DM$ and can be written in terms of partial wave expansion as $ \langle \sigma v \rangle = a +b v^2$. Numerical solution of the the Boltzmann equation gives \cite{Kolb:1990vq,Scherrer:1985zt}
\begin{equation}
\Omega_{\rm DM} h^2 \approx \frac{1.04 \times 10^9 x_F}{M_{\text{Pl}} \sqrt{g_*} (a+3b/x_F)}\,,
\end{equation}
where, $x_F = M_{\rm DM}/T_F$, $T_F$ is the freeze-out temperature, $M_{\rm DM}$ is the mass of dark matter, $g_*$ is the total relativistic degrees of freedom at the time of freeze-out and and $M_{\text{Pl}} \approx 1.22 \times 10^{19}$ GeV is the Planck mass. DM with electroweak scale mass freeze out at temperature approximately in the range $x_F \approx 20-30$. Where, $x_F$ can be calculated from the relation below 
\begin{equation}
x_F = \ln \frac{0.038gM_{\text{Pl}}M_{\rm DM}\langle \sigma v \rangle}{g_*^{1/2}x_F^{1/2}}\,,
\label{xf}
\end{equation}
\begin{figure}
\includegraphics[width=38mm,height=30mm]{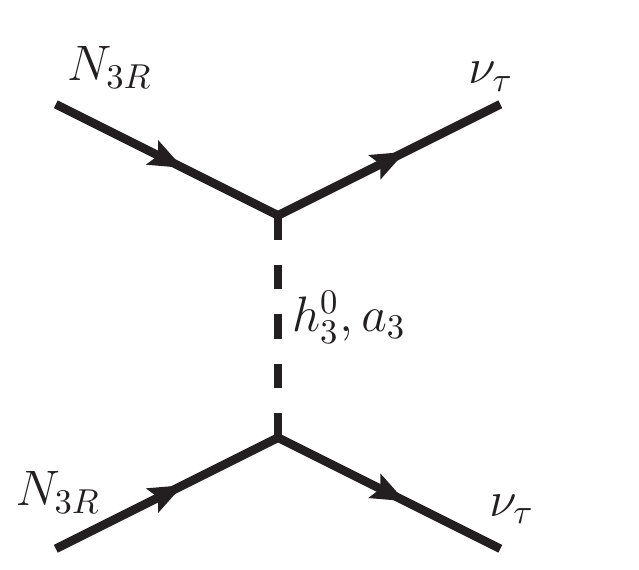}
\includegraphics[width=38mm,height=30mm]{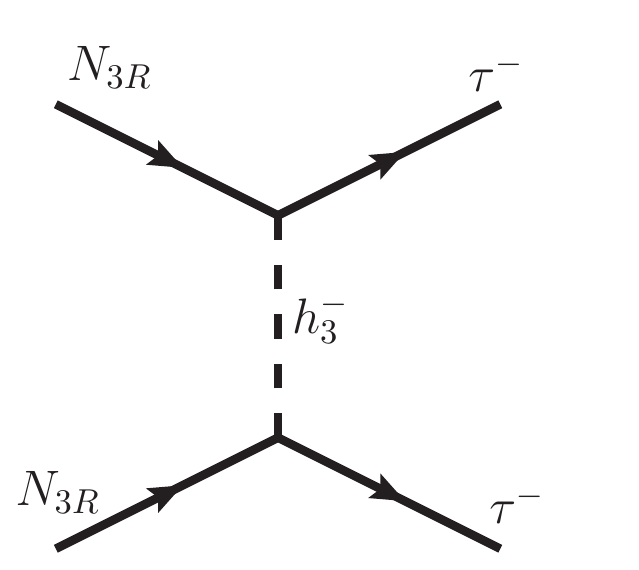}
\includegraphics[width=38mm,height=30mm]{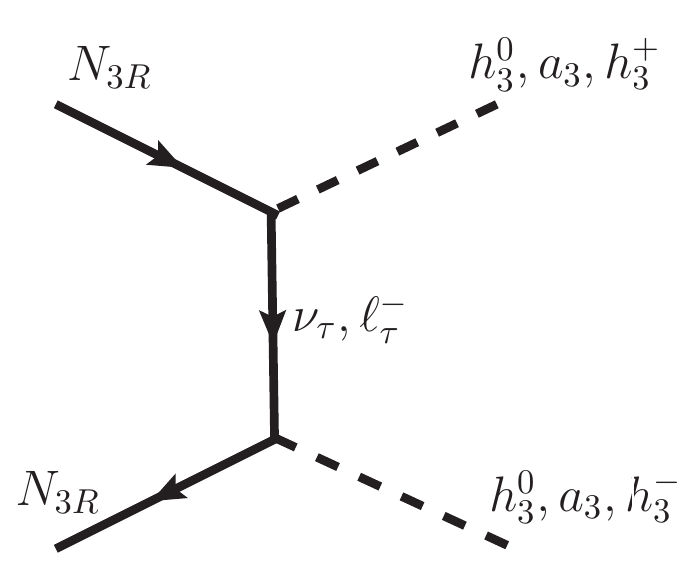}
\caption{t-channel annihilation of DM.}
\end{figure}
\begin{figure}
\includegraphics[width=70mm,height=54mm]{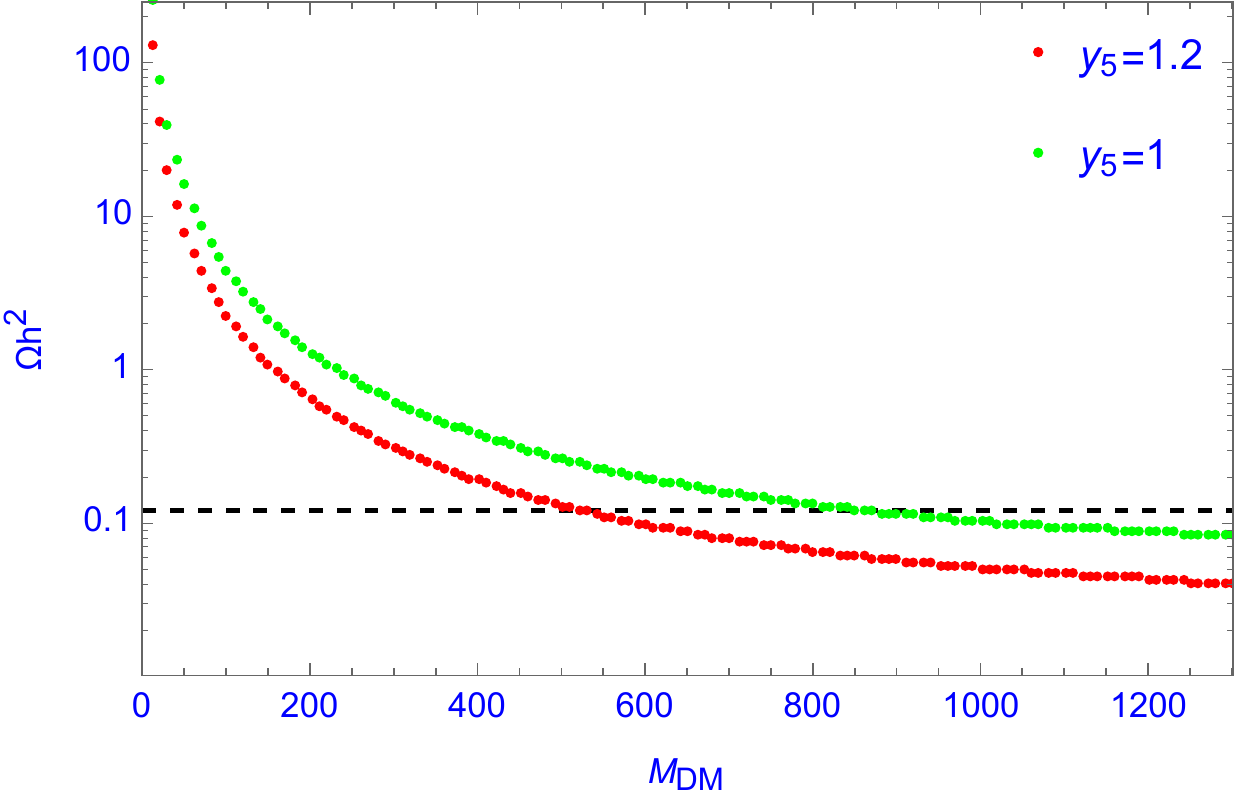}
\caption{Variation of dark matter mass with the relic density is shown in this figure. Here, the black dashed line represents the observed relic density as the Planck data \cite{Ade:2015xua} .}\label{DMrelic}
\end{figure}
As the Majorana DM has only interaction with the third generation lepton the corresponding Yukawa coupling plays an important role in the relic density perspective. The t-channel interactions mediated by the inert scalar and third generation leptons dominantly contribute to the relic density. In Fig.\ref{DMrelic}, one can infer that the correct relic can be achieved with a large Yukawa coupling of order 1 with a DM mass ($M_{DM}$, which is defined as $M_{3R}$ in the previous sections) more than 300 GeV.
\subsection{Loop Level Direct Searches}
\begin{figure}
\begin{center}
\includegraphics[height=40mm,width=55mm]{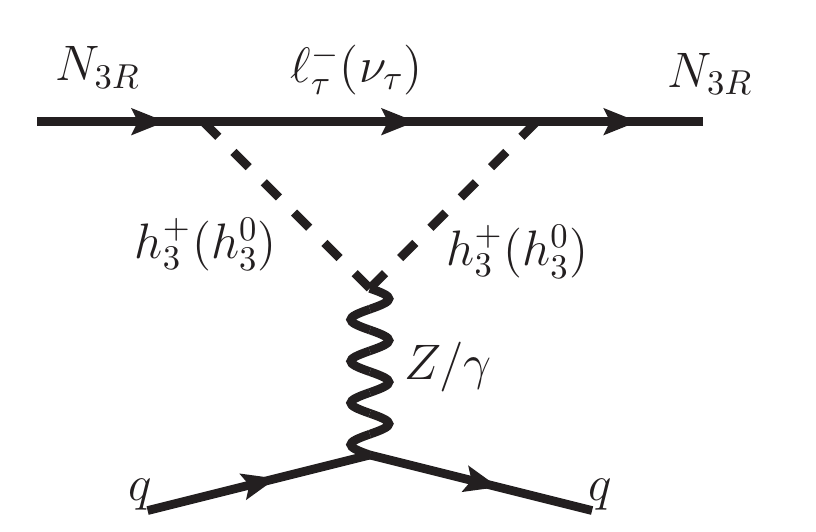} \hspace{6mm}
\includegraphics[height=40mm,width=58mm]{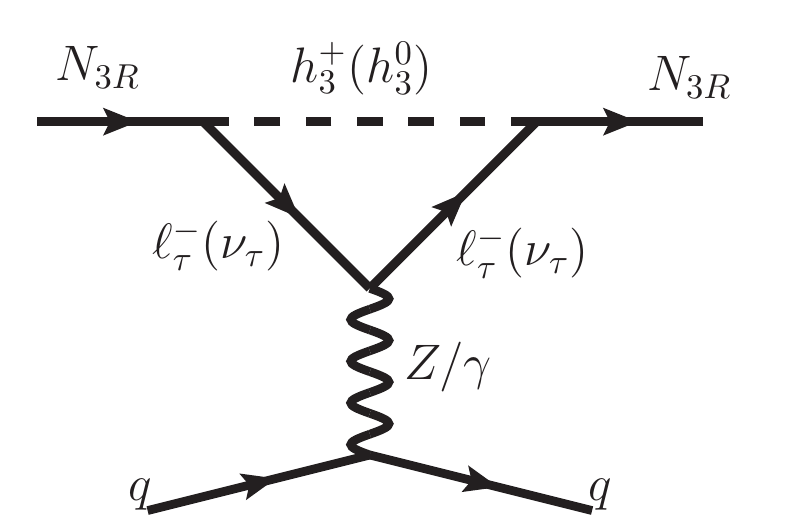}
\caption{Loop level Direct detection diagrams of the Majorana DM.}
 \end{center}
\end{figure}
Since the only interaction of the Majorana DM with the SM particles is through the Yukawa coupling, tree level direct detection process is not allowed in this model. But still the one loop effective interaction with the nucleus is possible through the availed couplings of DM. As the mixing of the $Z_2$ odd scalar with SM Higgs is not possible, the gauge boson mediated direct detection processes dominates in this framework. Mediation of Z boson leads to an effective axial vector interaction $\chi_q \overline{N_{3R}} \gamma_\mu \gamma_5 N_{3R} \bar{q} \gamma^\mu \gamma^5 q$ \cite{Ibarra:2016dlb}. Here,
\begin{equation}
\chi_q=\frac{y^2_5 a_q}{32 \pi^2 M^2_Z} \left[(g_l+a_l) F \left(\frac{M^2_{\rm DM}}{M^2_{h^+_3}}\right) + (g_\nu+a_\nu) F \left(\frac{M^2_{\rm DM}}{M^2_{h^0_3}}\right)\right].
\end{equation}
With $g_l=-\frac{g}{2\cos{\theta_w}}(\frac{1}{2}-2\sin^2{\theta_w})$, $a_l=-\frac{g}{4\cos{\theta_w}}$, $g_\nu=a_\nu=\frac{g}{4\cos{\theta_w}}$, $a_q=\frac{1}{2}(-\frac{1}{2})$ for $q=u,c,t (d,s,b)$. Where, g and $\theta_w$ are the gauge coupling and Weinberg mixing angle respectively. The loop function $F(x)$ is given by
\begin{eqnarray}
F(x)=−1+\frac{2(x+(1 - x){\rm ln}(1- x))}{x^2},
\end{eqnarray}
\begin{figure}
\begin{center}
\includegraphics[height=52mm,width=72mm]{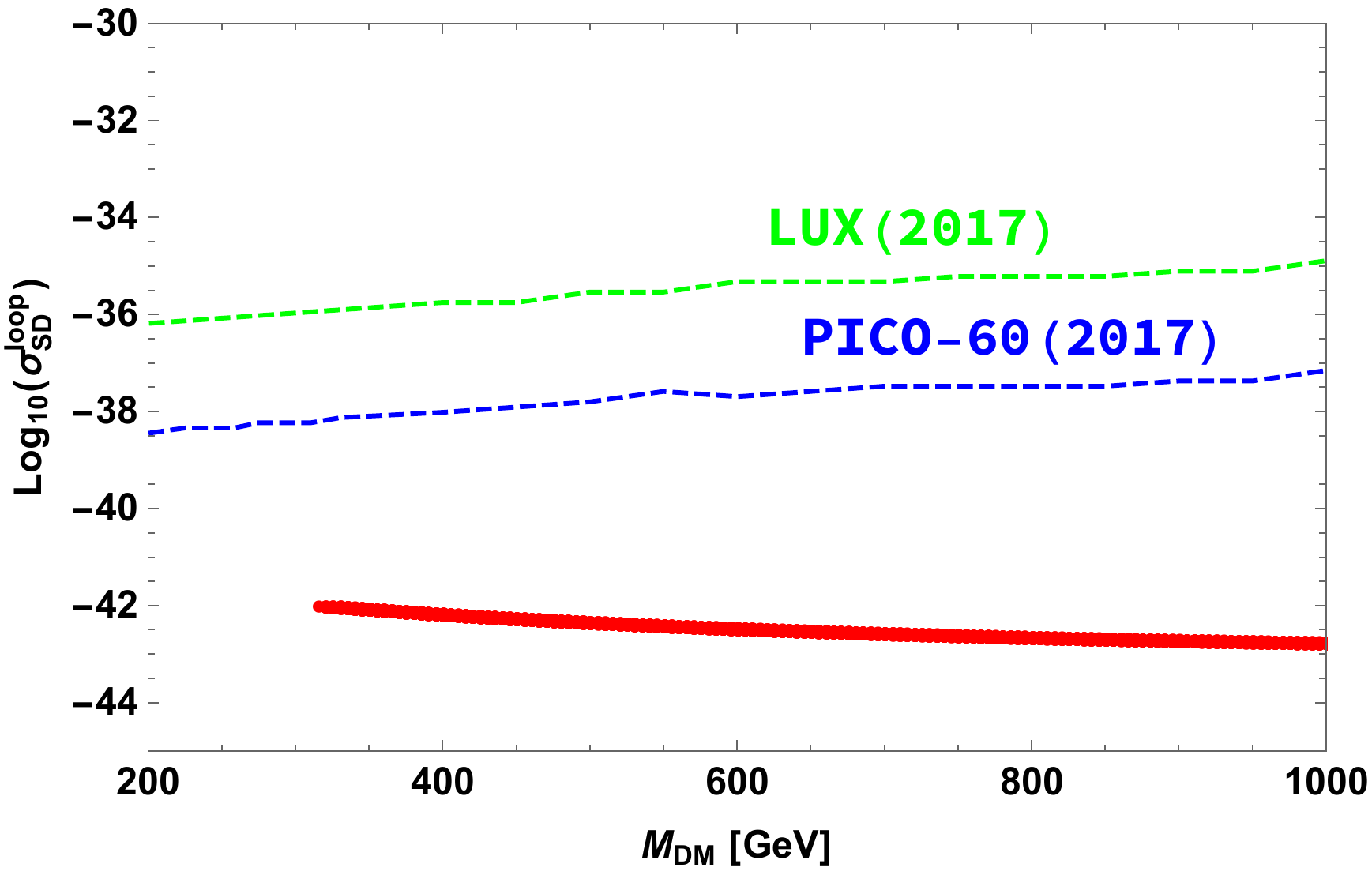}
\includegraphics[width=72mm,height=52mm]{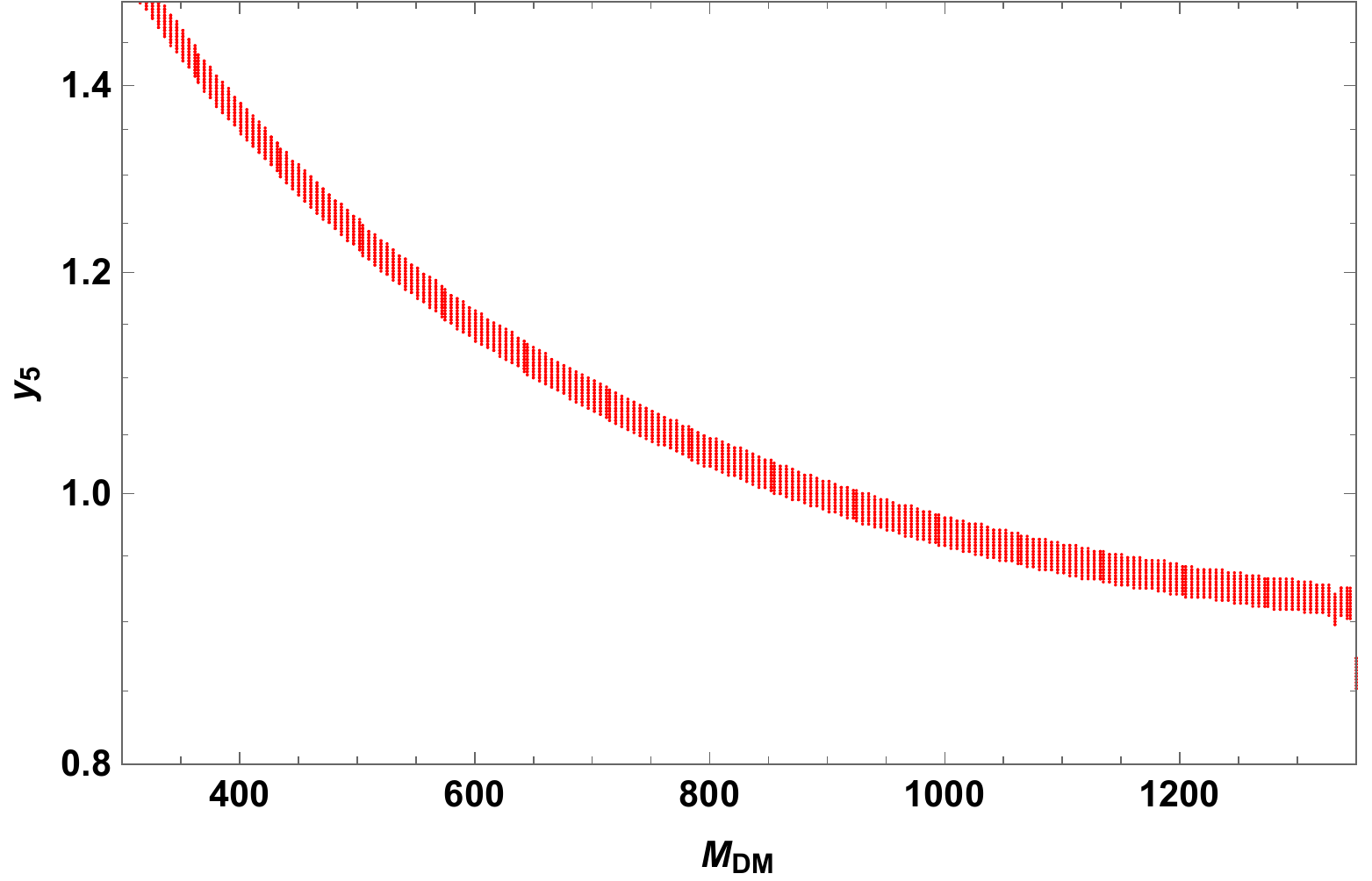}
\caption{Left panel represents the allowed parameter space for the DM mass as per the bound on spin dependent direct detection cross section from PICO-60\cite{Amole:2017dex} and LUX\cite{Akerib:2017kat} data. Here, the right panel shows the parameter space for DM mass and the corresponding Yukawa coupling, allowed by the observed $3\sigma$ value of relic density.}
 \end{center}
\end{figure}
 The spin dependent cross section of the nucleon is given by
\begin{equation}
\sigma_{SD}=\frac{16}{\pi} \frac{M^2_{\rm DM} + m^2_N}{(M_{\rm DM} + m_N)^2} J_N (J_N + 1) \chi^2_T.
\end{equation}
Here, $m_N$ and $J_N$ are the mass and spin of the nucleon respectively and $\chi_{T}=\sum_{q=u,d,s} \Delta_q \chi_q$. Where, $\Delta_q's$ are quark spin functions and are measured as $\Delta_u=0.842$,~ $\Delta_d=-0.427$,~$\Delta_s=-0.085$ \cite{Airapetian:2006vy}.
\section{Comment on Lepton Flavor Violation}
\begin{figure}
\includegraphics[height=32mm,width=52mm]{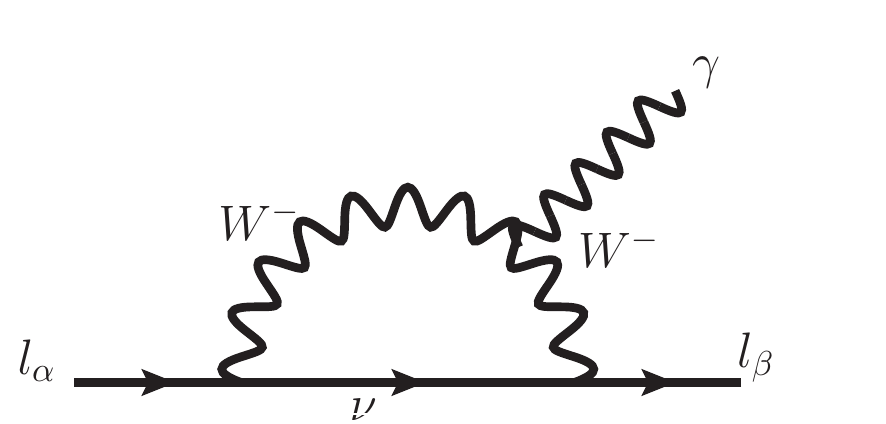}
\includegraphics[height=32mm,width=52mm]{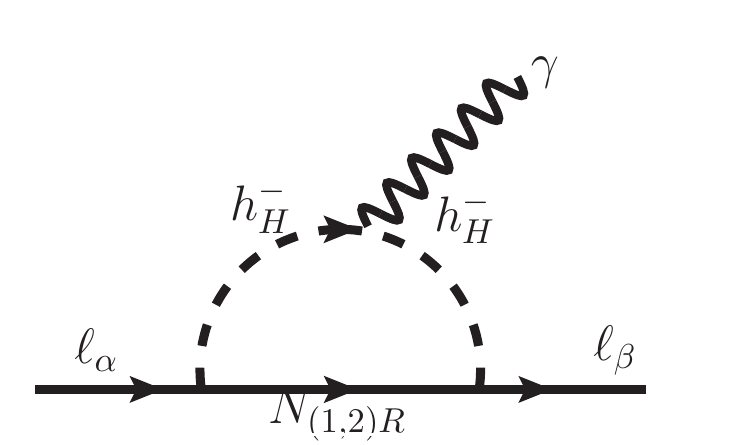}
\caption{Feynman diagrams represent the Lepton flavor violating rare decay ($\mu\rightarrow e\gamma$) in one loop.}\label{lfvfeyn}
\end{figure}
Lepton flavor violating decay processes have achieved a decent attention in current times \cite{Mihara:2013zna}-\cite{Dev:2017ftk}. With the efforts of many experiments to search for these rare signals, few of them have provided a stringent upper limits on these decay modes. In this framework, $\mu \rightarrow e\gamma$ decay process seems to be important, which is measured with less background from experimental perspective. The observed upper bound on the branching of this decay is Br$(\mu\rightarrow e\gamma)<4.2\times 10^{-13}$ from MEG collaboration \cite{TheMEG:2016wtm}. In this context, we can have additional contribution to this decay $l_\alpha\rightarrow l_\beta \gamma$ with TeV scale right handed neutrinos and Higgs. The branching ratio for this decay is given by \cite{Chekkal:2017eka}
\begin{figure}
\includegraphics[height=50mm,width=70mm]{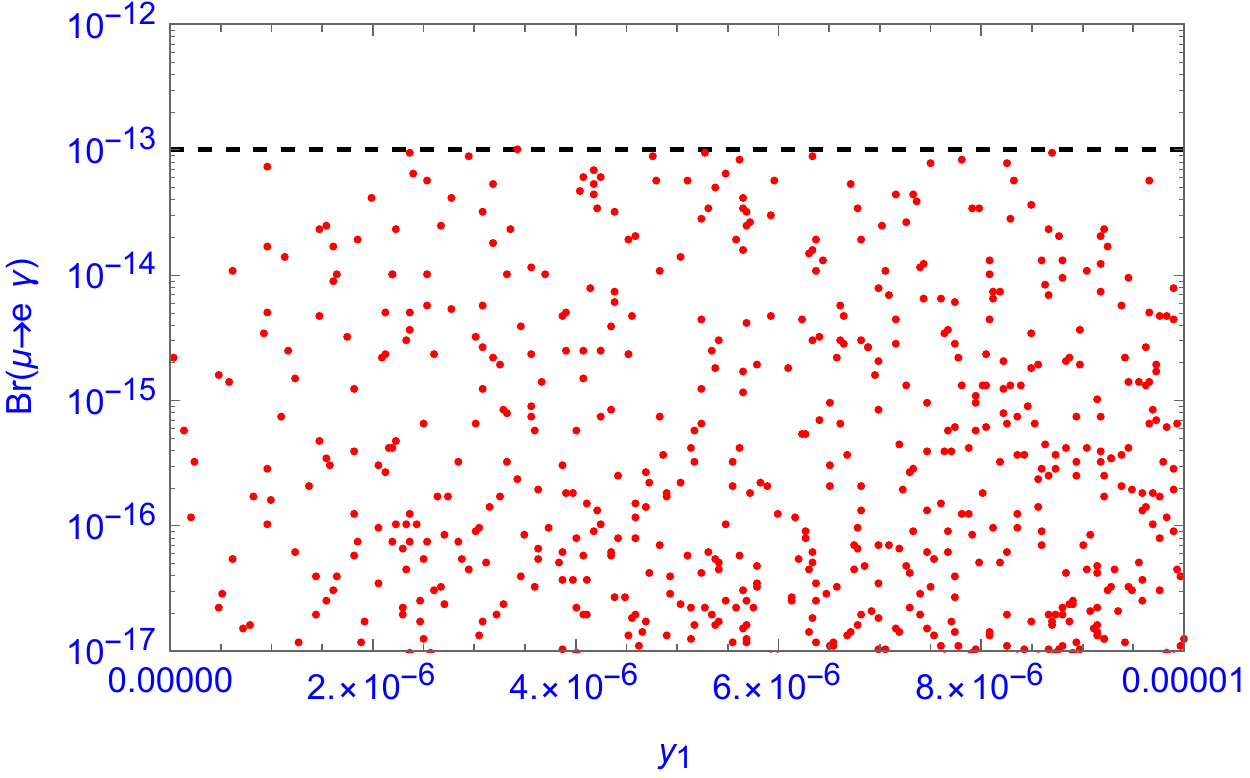}
\includegraphics[height=50mm,width=70mm]{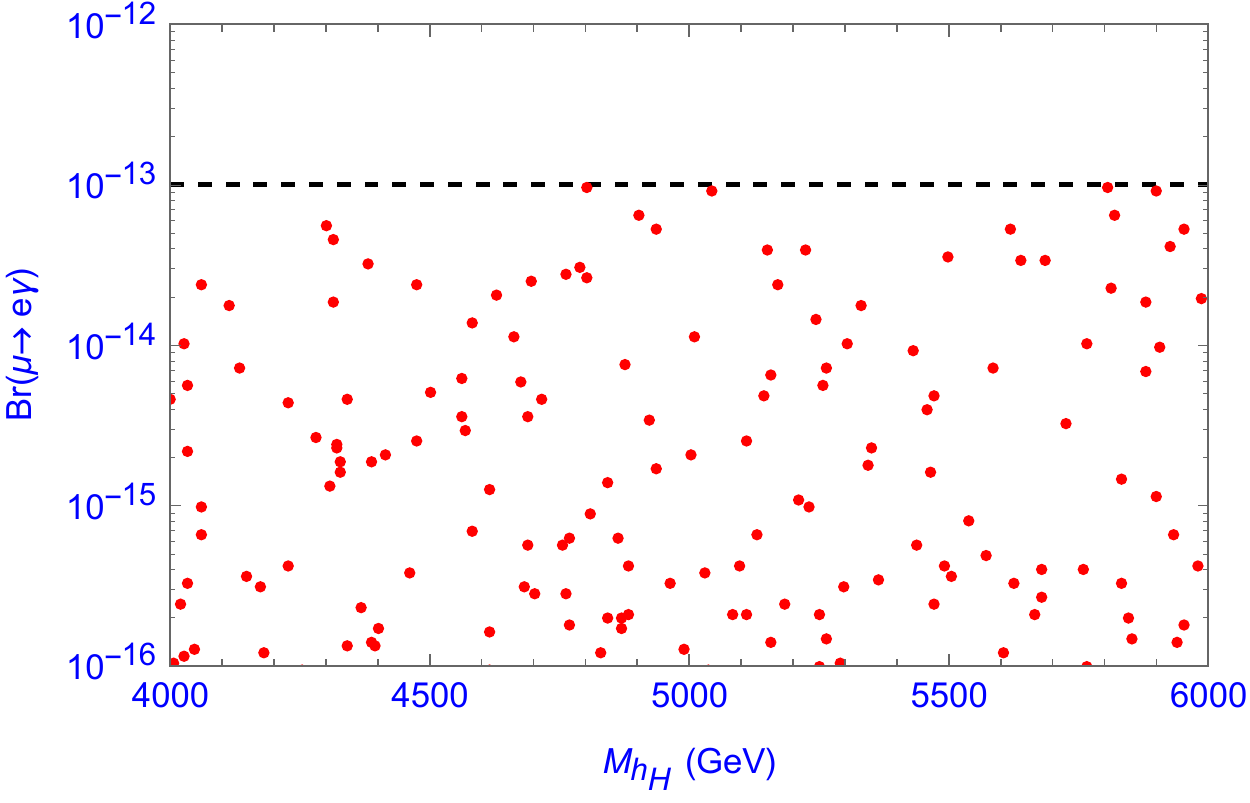}
\caption{Left panel represents the variation of Yukawa coupling and heavy Higgs mass with the branching of $\mu\rightarrow e\gamma$ respectively.}\label{lfv}
\end{figure}
\begin{equation}
Br(l_\alpha \rightarrow l_\beta \gamma)=\frac{3(4 \pi)^3 \alpha}{4 {G_F}^2}|A_D|^2\times Br(l_\alpha \rightarrow l_\beta \nu_\alpha \bar{\nu_{\beta}}).
\end{equation}
where, $G_F\approx 10^{-5}{\rm GeV}^{-2}$ is the Fermi coupling constant and $\alpha$ is the electromagnetic fine structure constant. $A_D$ is the dipole contribution, which is given by
\begin{equation}
A_D=\sum_i \frac{(Y^\nu_{H})_{\alpha i}(Y^{\nu \star}_{H})_{\beta i}f(x)}{2(4\pi)^2 M^2_{h_H}}.
\end{equation} 
Here, $Y^\nu_{H}$ and $M_{h_H}$ are the Yukawa coupling matrix \cite{Mishra:2019sye} and mass corresponds to the heavy Higgs. $f(x)$ is the loop function, with $x=\frac{M^2_{iR}}{M^2_{h_H}}$,~$i=1,2$, which is given by
\begin{equation}
f(x)=\frac{1-6x+3x^2+2x^3-6x^2 {\rm log}x}{6(1-x)^4}.
\end{equation} 
A common parameter space for the Yukawa coupling of two Heavy right-handed neutrinos from the neutrino mass is obtained, which satisfies the LFV constraints in one loop level flavor violating rare decay. To have a consistent LFV data one can infer from Fig.\ref{lfv} that the heavy Higgs mass should be of similar order with the right fermion masses.
\section{Summary}
In this article, an attempt has been made to explain neutrino mixing and dark matter phenomenology with a simplest permutation symmetry $S_3$ extension of standard model. In the current scenario, the specific structure of the neutrino mass matrix in tree level leaves one of the active neutrino to be massless. Hence like the scotogenic model, the radiative mass term for the mentioned neutrino is generated in one loop level, where, the loop is mediated by the odd particles. We constrained various model parameters as per the current $3\sigma$ observation of the neutrino oscillation. This model naively predicts a nonzero $\theta_{13}$, which is experimentally evidenced. Apart from the neutrino mixing, this model includes a Majorana dark matter candidate, which is stabilized by the $Z_2$ symmetry. The dark matter satisfies a correct relic as per the $3\sigma$ limit of Planck data, for a large Yukwa coupling of $\mathcal{O}(1)$ and a small $\lambda_7$, which retains the compatibility of observed neutrino oscillation data. As the Majorana fermion does not directly interact with the standard model quarks, one loop direct detection is discussed, which lies under the allowed bounds of LUX(2017) and PICO-60(2017). On the other hand, the TeV scale right-handed neutrinos and the heavy Higgs opens an option for the lepton flavor violating decay constraints from $\mu\rightarrow e \gamma$. Therefore this model seems to be interesting with a rich phenomenology to explain neutrino mass, dark matter and Lepton flavor violation constraints simultaneously and such a low scale new particles opts a future direction for the collider experiments.

I acknowledge DST Inspire for its financial support. I am thankful to Prof. Anjan Giri for his useful guidance and also acknowledge Nimmala Narendra and Dr. Shivaramakrishna Singirala for their help and discussions towards this work.

%\cite{Ibarra:2016dlb}

\end{document}